\documentclass[%
 reprint,
superscriptaddress,
showpacs,
 amsmath,amssymb,
 aps,
floatfix,
]{revtex4-1}
\usepackage{graphicx,xcolor,wrapfig}
\usepackage{dcolumn}
\usepackage{bm}

\usepackage[colorlinks=true,citecolor=blue,linkcolor=blue]{hyperref}

\begin{document}


\title{
Anomalous suppression of photo-induced in-gap weight in the optical conductivity of a two-leg Hubbard ladder}

\author{Takami Tohyama}
\email{tohyama@rs.tus.ac.jp}
\affiliation{Department of Applied Physics, Tokyo University of Science, Tokyo 125-8585, Japan}

\author{Kazuya Shinjo}
\email{kazuya.shinjo@riken.jp}
\affiliation{Computational Quantum Matter Research Team, RIKEN Center for Emergent Matter Science (CEMS), Wako, Saitama 351-0198, Japan}

\author{Shigetoshi Sota}
\affiliation{Computational Materials Science Research Team, RIKEN Center for Computational Science (R-CCS), Kobe, Hyogo 650-0047, Japan}
\affiliation{Quantum Computational Science Research Team, RIKEN Center for Quantum Computing (RQC), Wako, Saitama 351-0198, Japan}

\author{Seiji Yunoki}
\affiliation{Computational Quantum Matter Research Team, RIKEN Center for Emergent Matter Science (CEMS), Wako, Saitama 351-0198, Japan}
\affiliation{Computational Materials Science Research Team, RIKEN Center for Computational Science (R-CCS), Kobe, Hyogo 650-0047, Japan}
\affiliation{Quantum Computational Science Research Team, RIKEN Center for Quantum Computing (RQC), Wako, Saitama 351-0198, Japan}
\affiliation{Computational Condensed Matter Physics Laboratory, RIKEN Cluster for Pioneering Research (CPR), Saitama 351-0198, Japan}

\date{\today}
             

\begin{abstract}
Photoinduced nonequilibrium states in the Mott insulators reflect the fundamental nature of competition between itinerancy and localization of the charge degrees of freedom. The spin degrees of freedom will also contribute to the competition in a different manner depending on lattice geometry. We investigate pulse-excited optical responses of a half-filled two-leg Hubbard ladder and compare them with those of a one-dimensional extended Hubbard chain.  Calculating the time-dependent optical conductivity, we find that strong monocycle pulse inducing quantum tunneling gives rise to anomalous suppression of photo-induced in-gap weight, leading to negative weight. This is in contrast to finite positive weight in the Hubbard chain. The origin of this anomalous behavior in the two-leg ladder is attributed to photoinduced localized exciton that reflects strong spin-singlet dimer correlation in the ground state.

\end{abstract}
\maketitle


\section{Introduction}
\label{Sec1}
The nonequilibrium-induced insulator-to-metal transition is a fundamental issue associated with competition between itinerancy and localization of charge degrees of freedom. Photoinduced insulator-to-metal transitions due to photon absorption have been observed in one-dimensional (1D) Mott insulators~\cite{Taguchi2000, Iwai2003, Okamoto2007, Al-Hassanieh2008, Wall2011}. Nonabsorbable terahertz photons with strong intensity have also been suggested to induce a metallic state~\cite{Liu2012, Yamakawa2017} via quantum tunneling~\cite{Oka2003, Oka2005, Oka2008, Oka2010, Eckstein2010, Oka2012}. Very recently, the emergence of glassy phase in  quantum tunneling has been proposed for a 1D Mott insulator through an insulator-to-glass transition when mono- and half-cycle pulses are applied~\cite{Shinjo2022}.

In contrast to well-studied 1D Mott insulators, photoinduced nonequilibrium properties in two-leg ladder Mott insulators have not been elucidated, although doped two-leg Mott insulators Sr$_{14-x}$Ca$_x$Cu$_{24}$O$_{41}$ have been examined experimentally~\cite{Fukaya2015} and theoretically~\cite{,Hashimoto2016,Shao2019}. A crucial difference in Mott insulators between 1D chain and two-leg ladder is spin excitation from the ground state: there is a spin gap in the ladder but gapless excitation in the chain. The spin gap in the ladder is qualitatively ascribed to an energy to break spin-singlet dimer predominately formed along the rung~\cite{Dagotto1996}. Because of strong spin dimer formation in the two-leg Mott insulator, dimer-dimer correlation is also strong as compared with the 1D Mott insulator. Under the presence of strong dimer-dimer correlation, photoexcited doublon and holon tend to form a localized exciton not to break spin dimers. Such localized exciton formation appears in optical absorption as a large peak at a high frequency slightly above the on-site Coulomb energy $U$ of a large $U$ Hubbard ladder~\cite{Shinjo2021}. In contrast, there is a small hump at $U$ in a 1D Hubbard chain~\cite{Gebhard1997,Essler2001,Jeckelmann2003}. Therefore, one can expect different photoinduced nonequilibrium properties in Mott insulators between 1D chain and two-leg ladder if the localized exciton contributes to photoexcitation. In particular, it is interesting to know whether photoinduced glassy dynamics proposed in a Hubbard chain~\cite{Shinjo2022} remains or not in a two-leg Hubbard ladder.

In this paper, we investigate pulse-excited states of a half-filled two-leg Hubbard ladder using the time-dependent density-matrix renormalization group (tDMRG) implemented by the Legendre polynomial~\cite{Shinjo2022} and time-dependent exact diagonalization (tED) based on the Lanczos technique~\cite{Lu2015}. We find that nonabsorbable monocycle pulse induces quantum tunneling as in the case of a 1D chain but photoinduced in-gap optical weight inside the Mott gap behaves differently from the 1D case with a glassy behavior. The spectral weight becomes negative if the monocycle pulse is strong enough. The origin of this anomalous behavior in the two-leg ladder is attributed to photoinduced localized exciton.

This paper is organized as follows. We introduce a two-leg Hubbard ladder and a method to calculate the time-dependent optical conductivity in Sec.~\ref{Sec2}. In Sec.~\ref{Sec3}, we investigate photoinduced nonequilibrium properties of the two-leg Hubbard ladder after applying a monocycle pump pulse. To understand the origin of photo-induced in-gap weight discussed in Sec.~\ref{Sec3}, we examine the time-dependent optical conductivity for a multicycle pulse in Sec.~\ref{Sec4}.  Finally, we give a summary of this work in Sec.~\ref{Sec5}.  Note that in this paper, we set the light velocity $c$, the elementary charge $e$, the Dirac constant $\hbar$, and the lattice constant to be 1.

\section{Model and method}
\label{Sec2}

A two-leg Hubbard ladder with a vector potential $A_\text{leg}(t)$ applied along the leg direction is modeled as 
\begin{align}\label{Hlad}
\mathcal{H}=&-t_\text{h} \sum_{i,\sigma} \left( e^{iA_\text{leg}(t)} \sum_\alpha c_{i,\alpha,\sigma}^\dag c_{i+1,\alpha,\sigma} + c_{i,1,\sigma}^\dag c_{i,2,\sigma} + \text{H.c.} \right) \nonumber \\
+& U\sum_{i,\alpha} n_{i,\alpha,\uparrow}n_{i,\alpha,\downarrow},
\end{align}
where $c_{i,\alpha,\sigma}^{\dag}$ is the creation operator of an electron with spin $\sigma (= \uparrow, \downarrow)$ at the $i$th rung of the $\alpha$ leg, $\alpha=1,2$; $n_{i,\alpha,\sigma}=c^\dagger_{i,\alpha,\sigma}c_{i,\alpha,\sigma}$; $t_\text{h}$ is the nearest-neighbor hopping; and $U$ is the on-site repulsive Coulomb interaction. Spatially homogeneous electric field $E_\text{leg}(t)=-\partial_{t}A_\text{leg}(t)$ is incorporated via the Peierls substitution in the hopping terms~\cite{Peierls1933}. In the following, we set $t_\text{h}$ and $t_\text{h}^{-1}$ as energy and time units, respectively.

We assume that pulses have time dependence determined by $A_\text{leg}(t)=A_\text{pump}(t)+A_\text{probe}(t)$ with $A_\text{pump}(t)=A_0 e^{-(t-t_0)^2/(2t_\mathrm{d}^2)} \cos\left[\Omega(t-t_0)\right]$ for a pump pulse and $A_\text{probe}(t)=A_0^\text{pr} e^{-\left(t-t_0^\text{pr}\right)^2/\left[2(t_\mathrm{d}^\text{pr})^2\right]} \cos \left[\Omega^\text{pr}(t-t_0^\text{pr})\right]$ for a probe pulse. The electric field amplitude $E_{0}$ of a pump pulse is obtained from $-\partial_{t}A_\text{pump}(t)$. We set $A_0^\text{pr}=0.001$, $\Omega^\text{pr}=10$, $t_\text{d}^\text{pr}=0.02$, and $t_{0}^\text{pr}=t_{0}+\tau$, where $\tau$ indicates the delay time between pump and probe pulses. We obtain time-dependent wave functions by the tDMRG implemented by the Legendre polynomial~\cite{Shinjo2021} employing open boundary conditions and keep $\chi=8000$ density-matrix eigenstates. We also use the tED based on the Lanczos technique~\cite{Lu2015}.

We obtain the optical conductivity in nonequilibrium~\cite{Lu2015}
\begin{equation}\label{sigma}
\sigma(\omega,\tau) = \frac{1}{L}\frac{j_\text{probe}(\omega,\tau) }{i(\omega +i\gamma)A_\text{probe}(\omega)},
\end{equation}
where $A_{\text{probe}}(\omega)$ and $j_\text{probe}(\omega,\tau)$ are the Fourier transform of $A_\text{probe}(t)$ and that of induced current $j_\text{probe}(t,\tau)$ $(t>t_{0}^\text{pr})$ due to a probe pulse along the leg direction, respectively, $L$ is the number of lattice sites, and $\gamma$ is a broadening factor.

\begin{figure}[t]
\includegraphics[width=0.45\textwidth]{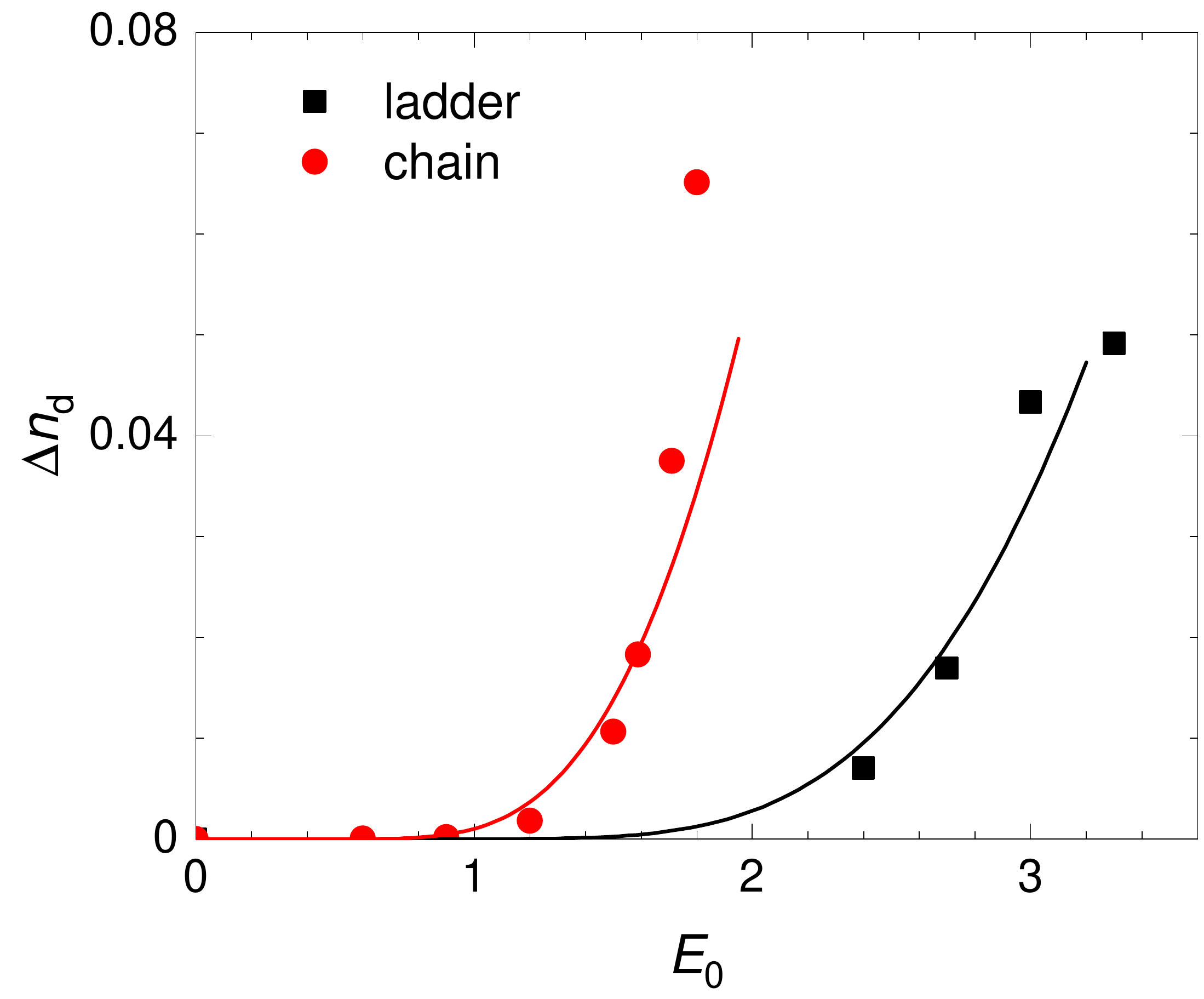}
\caption{Photoinduced doublon density $\Delta n_\text{d}$ of half-filled Hubbard lattices as a function of the strength of a monocycle pulse with $\Omega=0$ and $(t_\text{d},t_{0})=(2,10)$. The density is calculated at $t=22$. Black and red symbols are data for the $L=16\times 2$ two-leg ladder ($U=10$) and $L=32$ chain ($U=10$ and $V=3$), respectively. The black and red lines show a fitted curve using Eq.~(\ref{EDC}) with $E_\text{th}=4.0$ and $2.1$, respectively.}
\label{Fig1}
\end{figure}

\section{Monocycle pulse}
\label{Sec3}

In a one-dimensional Mott insulator described by a half-filled extended Hubbard chain, the change of doublon density $\Delta n_\text{d}$ induced by a monocycle electric pulse follows a threshold behavior~\cite{Shinjo2022} given by the form~\cite{Oka2012}
\begin{equation}\label{EDC}
\Delta n_\text{d} \propto E_{0}\exp \left(-\pi \frac{E_\text{th}}{E_{0}}\right),
\end{equation}
where $E_\text{th}$ is the threshold field. To clarify whether a half-filled two-leg Hubbard ladder shows a threshold behavior, we apply an $\Omega=0$ monocycle pulse with $(t_\text{d},t_{0})=(2,10)$ to an $L=16\times 2$ Hubbard ladder, and perform tDMRG calculations of $\Delta n_\text{d}$ defined as $\Delta n_\text{d}=\bigl[ \langle N \rangle_{t=22} - \langle N\rangle_{t<0} \bigr]/L$, where $\langle N \rangle_{t=22}$ is an expectation value of $N=\sum_{i,\alpha}n_{i,\alpha,\uparrow}n_{i,\alpha,\downarrow}$ just before a probe pulse with $t_{0}^\text{pr}=22$ is applied and $\langle N \rangle_{t<0}$ is for the ground state.

Figure~\ref{Fig1} shows that $\Delta n_\text{d}$ for a two-leg Hubbard ladder (black squares) also follows a threshold behavior as evidenced by a fit using Eq.~(\ref{EDC}) (black line).  Here, we assume that  Eq.~(\ref{EDC}) obtained for a one-dimensional Mott insulator is applicable for a two-leg ladder, since the electric field is applied only along the leg direction. From the fit, we find that an estimated value of $E_\text{th}=4.0$ for the two-leg ladder is larger than $E_\text{th}=2.1$ for a half-filled extended Hubbard chain with $U=10$ and the nearest-neighbor Coulomb interaction $V=3$ (red circles and red line)~\cite{Shinjo2022}. We note that almost double difference in $E_\text{th}$ cannot be explained by only 1.25 times difference in the Mott-gap value, $E_\text{gap}=7.5$ for the two-leg ladder and $E_\text{gap}=6$ for the chain~\cite{ratio}. The correlation length between photoinduced doublon and holon can be estimated from a relation $\xi \simeq E_\text{gap}/(2E_\text{th})$~\cite{Oka2012}. We find that $\xi\sim 0.91$ for the two-leg ladder, which is shorter than $\xi\sim 1.5$ for the chain. The shorter correlation length might be related to spatial extension of a doublon-holon pair on a rung, which effectively reduces an extension toward the leg direction. This is consistent with a view that photoexcited doublon and holon tend to form a localized exciton under the presence of strong dimer-dimer correlation, as mentioned in the introduction.

\begin{figure}[t]
\includegraphics[width=0.45\textwidth]{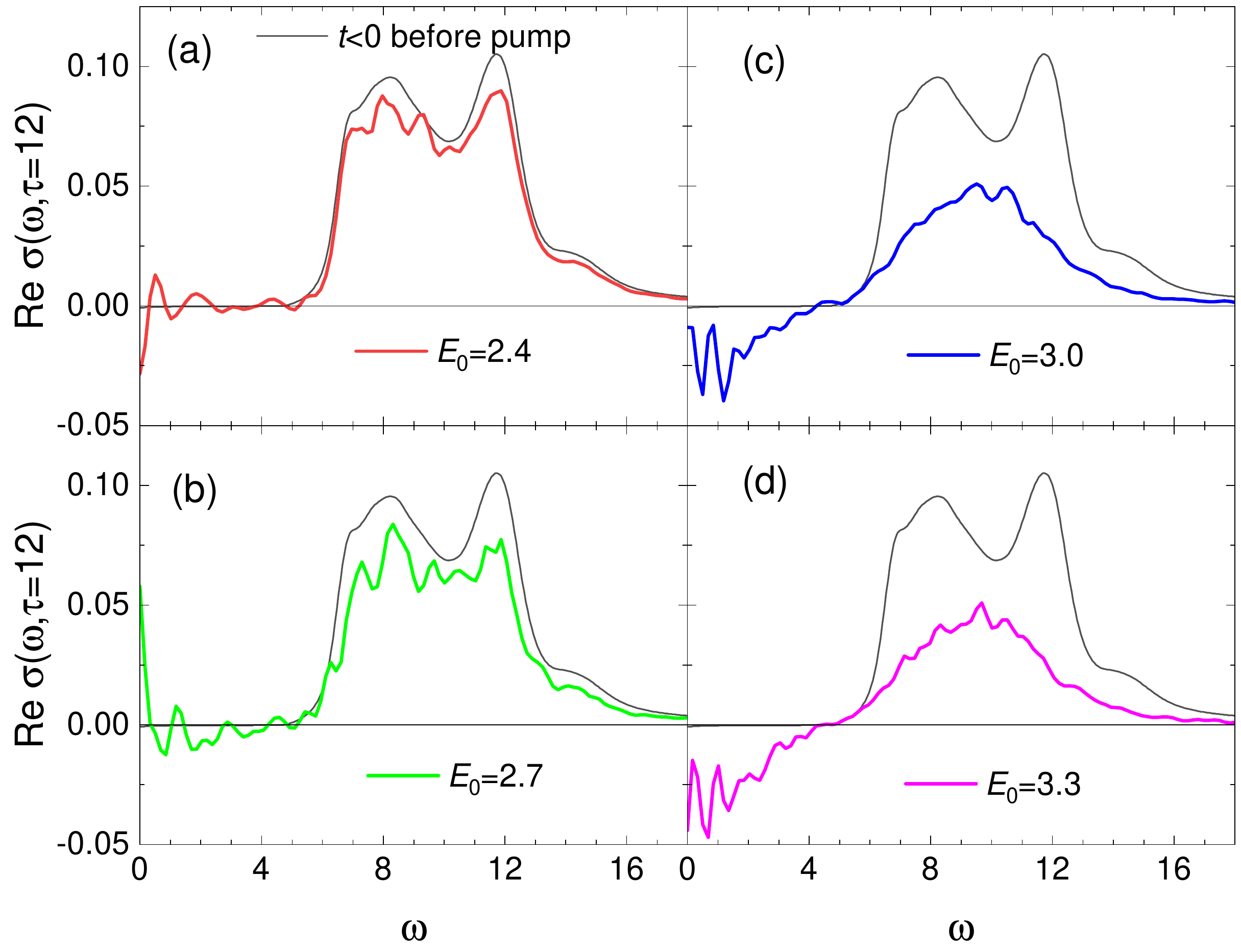}
\caption{$\text{Re}\,\sigma(\omega,\tau=12)$ of the half-filled $L=16\times 2$ two-leg Hubbard ladder ($U=10$) excited by the $\Omega=0$ monocycle pulse with $(t_\text{d},t_{0})=(2,10)$. (a) $E_{0}=2.4$, (b) $E_{0}=2.7$, (c) $E_{0}=3.0$, and (d) $E_{0}=3.3$. Black thin line on each panel is the case $t<0$ before pumping. The broadening factor is $\gamma=0.4$.}
\label{Fig2}
\end{figure}

We show $\text{Re}\,\sigma(\omega,\tau)$ excited by a monocycle pulse with $(\Omega,t_\text{d})=(0,2)$ for various $E_{0}$ in Fig.~\ref{Fig2}. The thin solid line in each panel represents $\text{Re}\,\sigma(\omega,\tau<0)$ before pumping. The equilibrium optical conductivity exhibits a broad two-peak structure above the Mott gap. A high energy peak at $\omega=12$ is due to a localized exciton originated from dimer-dimer correlation in the ground state~\cite{Shinjo2021}. We call this peak the localized exciton peak with energy $\omega_\text{L}$. At $E_0=2.4$ shown in Fig.~\ref{Fig2}(a), the spectral weight of the two-peak structure is reduced almost uniformly and small weight below the Mott gap emerges due to the presence of photoinduced doublon and holon. At $E_0=2.7$, more suppression of the broad peak as well as weight transfer to low-energy Drude components is seen in Fig.~\ref{Fig2}(b). With further increasing $E_0$, the localized exciton peak is completely suppressed and negative weight inside the Mott gap becomes significant as shown in Figs.~\ref{Fig2}(c) and \ref{Fig2}(d). Such negative weight ranging whole in-gap region has not been observed in a one-dimensional extended Hubbard chain~\cite{Shinjo2022}, where in-gap weight at finite frequency increases with increasing $E_0$. Therefore, anomalous suppression with negative weight with increasing $E_0$ is characteristic of a photoexcited two-leg ladder.The anomalous suppression seems to be related to the disappearance of the localized exciton peak. We will discuss this point in Sec.~\ref{Sec4}.

\begin{figure}[t]
\includegraphics[width=0.4\textwidth]{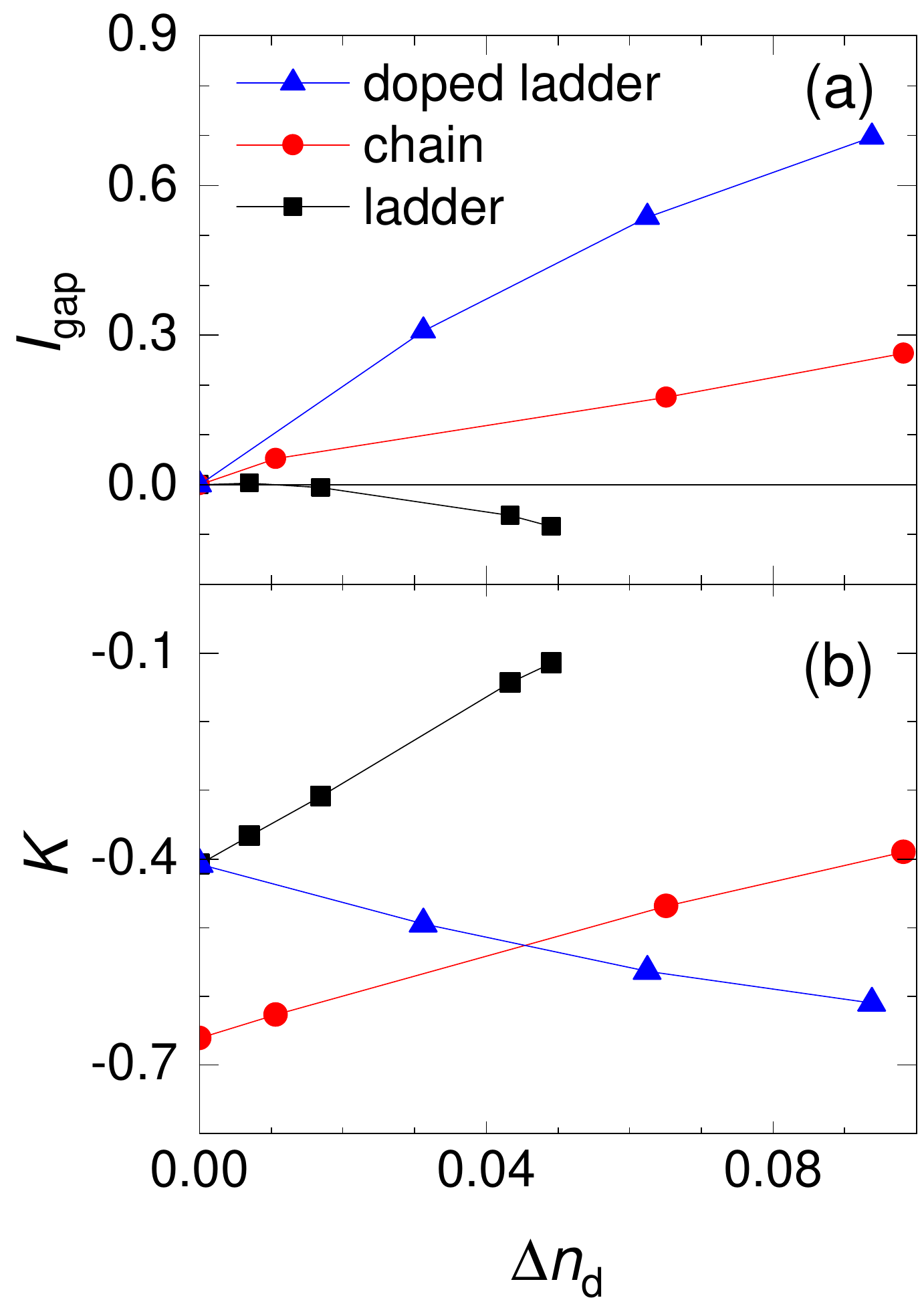}
\caption{(a) Integrated spectral weight $I_\text{gap}$ of the real part of optical conductivity inside the Mott gap ($\omega<5$) and (b) the kinetic energy $K$ along the leg direction, as a function of $\Delta n_\text{d}$. Black and red symbols are data for the half-filled $L=16\times 2$ two-leg Hubbard ladder ($U=10$) and the $L=32$ Hubbard chain ($U=10$ and $V=3$), respectively. Blue triangles are the data for the electron-doped $L=16\times 2$ two-leg Hubbard ladder ($U=10$). The solid line is a guide to the eye.}
\label{Fig3}
\end{figure}

Integrated photoinduced in-gap weight defined as $I_\text{gap}=\int_0^5\text{Re}\,\sigma (\omega,\tau=12)d\omega$ is shown in Fig.~\ref{Fig3}(a) as a function of $\Delta n_\text{d}$ for both the two-leg ladder (black squares) and chain (red  circles). For comparison, in-gap weight for an electron-doped two-leg Hubbard ladder (blue triangles) is also shown, where $\Delta n_\text{d}$ is set to be a half of electron-carrier density to compare with carrier density of pulse-excited systems where the same number of doublon and holon are excited. While $I_\text{gap}$ for the chain and doped ladder increases with $\Delta n_\text{d}$, $I_\text{gap}$ for the ladder changes from almost zero to negative values. 

To understand this anomalous behavior of in-gap weight, we calculate the kinetic energy along the leg direction at $t=22$ ($\tau=12$): 
\begin{equation}\label{Kleg}
K= -\frac{t_\text{h}}{L} \sum_{i,\alpha,\sigma} \langle c_{i,\alpha,\sigma}^\dag c_{i+1,\alpha,\sigma} + \text{H.c.}  \rangle
\end{equation}
whose absolute value is expected to be proportional to total integrated weight of $\text{Re}\,\sigma(\omega,\tau=12)$ according to the optical sum rule in the equilibrium case~\cite{Maldague1977}. Figure~\ref{Fig3}(b) shows $K$ as a function of $\Delta n_\text{d}$. For the electron-doped ladder, $\left|K\right|$ increases with $\Delta n_\text{d}$ as expected from doped Mott insulator~\cite{Dagotto1994}. On the other hand, $\left|K\right|$ for both the photodoped ladder and chain decreases, indicating the decrease of total integrated weight of optical conductivity. In the chain, the decrease of $\left|K\right|$ is also related to the emergence of glassy dynamics when the strength of a monocycle pulse increases~\cite{Shinjo2022}.  Interestingly, $\left|K\right|$ for the two-leg ladder approaches zero more quickly than that for the chain. This small $\left|K\right|$ means a strong localization effect of photoinduced doublon and holon in the two-leg ladder. Since spectral weight above $E_\text{gap}$ remains significant in $\text{Re}\,\sigma(\omega,\tau=12)$, the small $\left|K\right|$ inevitably leads to negative in-gap weight.

\begin{figure}[t]
\includegraphics[width=0.45\textwidth]{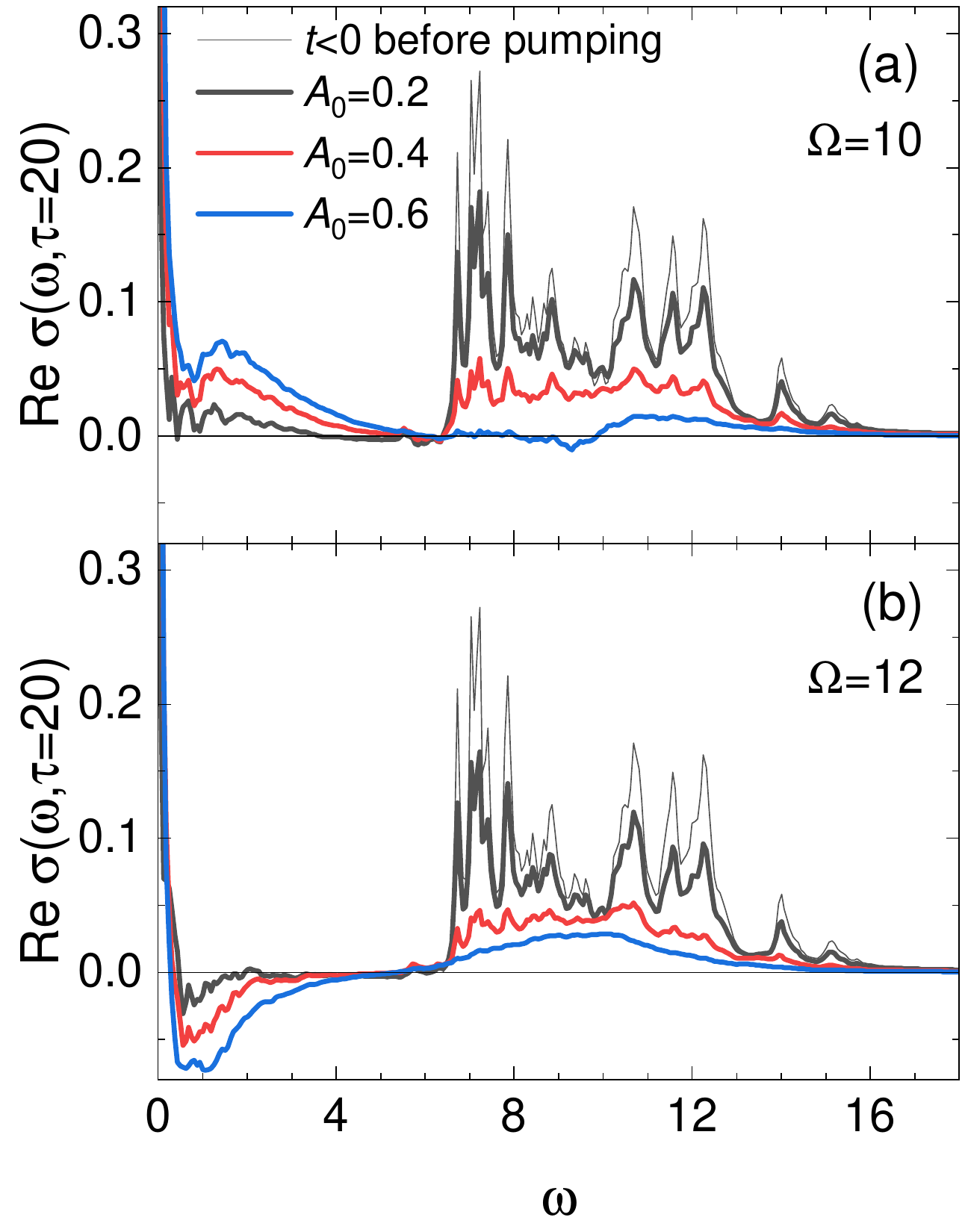}
\caption{$\text{Re}\,\sigma(\omega,\tau=20)$ of the half-filled $L=8\times 2$ two-leg Hubbard ladder ($U=10$) excited by (a) $\Omega=10$ and (b)$\Omega=12$ multicycle pulses with $t_\text{d}=1$. Black solid, red solid, and blue solid lines are for $A_0=0.2$, $0.4$, and $0.6$, respectively. Black thin line on each panel is the case for $t<0$ before pumping.  The broadening factor $\gamma=0.07$.}
\label{Fig4}
\end{figure}

\section{Multicycle pulse}
\label{Sec4}
It is necessary to clarify whether such strong localization effect in the two-leg Hubbard ladder appears only for a monocycle pump pulse or not. A hint would lie in the strong suppression of the localized exciton peak at $\omega_\text{L}=12$ in Figs.~\ref{Fig2}(c) and \ref{Fig2}(d). It is expected that negative in-gap weight will appear if one strongly excites states corresponding to the peak. To confirm this speculation, we perform the tED calculation of $\text{Re}\,\sigma(\omega,\tau)$ at $\tau=20$ using a multicycle pulse that excites states above the Mott gap~\cite{Size}. We apply the pulse with $(t_\text{0},t_\text{d})=(5,1)$ to a $8\times 2$ Hubbard ladder with $U=10$. To compare resonant and off-resonant cases with the localized exciton peak around $\omega_\text{L} \sim12$ shown in Fig.~\ref{Fig4}, we consider two pumping frequencies: $\Omega=10$ [Fig.~\ref{Fig4}(a)] and $\Omega=12$ [Fig.~\ref{Fig4}(b)]. The two cases exhibit a clear difference inside the Mott gap: in-gap weight for $\Omega=10$ increases with increasing the strength of pump pulse $A_0$, while it decreases and becomes negative for $\Omega=12$.

If we excite states corresponding to the localized exciton peak by a pumping photon, the time-dependent wavefunction starts to have the component of a localized doublon-holon pair. As a result, the absolute value of kinetic energy along the leg direction $\left|K\right|$ is expected to be significantly reduced. This is the case for Fig.~\ref{Fig4}(b): $K=-0.45$ before pumping changes to $K=-0.07$ at $A_0=0.6$. Therefore, in-gap weight has to be negative to achieve a small value of total integrated weight (proportional to $\left|K\right|$) in the presence of a large amount of spectral weight above the Mott gap.  This mechanism does not hold when we excite states below the localized exciton peak as shown in Fig.~\ref{Fig4}(a). The suppression of the localized exciton peak and emergence of negative in-gap weight are qualitatively similar to the case of a monocycle pulse mentioned in Sec.~\ref{Sec3}. 
This correspondence clearly evidences the fact that the suppression of localized exciton peak induces photoinduced negative in-gap weight. 

We note that negative spectral weight is not necessarily a common phenomenon when spin-singlet dimer correlation is strong in the ground state. In a large $U$ Hubbard chain with second-nearest-neighbor hopping $t'=0.7$, an effective low-energy spin model is expected to the $J_1$-$J_2$ model with $J_2/J_1\sim 0.5$, which is known as the Majumdar-Ghosh model, whose ground state is a product of singlet-dimer pairs~\cite{Majumdar1969}, where $J_1$ ($J_2$) is the first (second) nearest-neighbor antiferromagnetic exchange interaction. In this Hubbard chain, a localized exciton peak emerges as for the two-leg Hubbard ladder. However, we did not find in-gap negative weight even if we selectively excite states corresponding to the localized exciton peak (not shown here). Therefore, the presence of spin dimer correlation is not enough for realizing negative in-gap weight.

From these results, we can take a following view of photoexcitation using a monocycle pulse for a two-leg Mott insulator.  The strong monocycle pulse creates photoinduced doublon-holon pairs but strong spin dimer-dimer correlation inherent in a two-leg ladder geometry prevents from extending the doublon-holon pairs. This results in the increase of localized doublon-holon components in the time-dependent wavefunction, inducing both the reduction of $\left|K\right|$ and the suppression of the localized exciton peak in the optical conductivity. Since a large amount of optical absorption still remains above the Mott gap, photoinduced in-gap spectral weight inevitably becomes negative due to the reduction of $\left|K\right|$ through the optical sum rule.

\section{Summary}
\label{Sec5}

	In summary, we have investigated $\text{Re}\,\sigma(\omega,\tau)$ of pulse-excited states of the half-filled two-leg Hubbard ladder using tDMRG and tED. We have found that strong monocycle pulse inducing quantum tunneling gives rise to anomalous suppression of photoinduced in-gap weight, leading to negative weight. This is in contrast to finite positive weight in an extended Hubbard chain. Examining multipulse pumping for states above the Mott gap, we have attributed the origin of this anomalous behavior of in-gap spectral weight to photoinduced localized exciton that reflects strong spin-singlet dimer correlation in the ground state. This contrasting behavior between the 1D chain and two-leg ladder Mott insulators can be confirmed if one applies a monocycle terahertz pulse to a 1D Mott insulator such as Sr$_2$CuO$_3$ and a two-leg ladder Mott insulator such as SrCu$_2$O$_3$~\cite{Hiroi1991} and La$_6$Ca$_8$Cu$_{24}$O$_{41}$~\cite{Carter1996}, increases pulse strength, and observes how in-gap weight evolves with the strength. Using typical parameters (lattice constant $\sim 0.4$~nm, and $t_\mathrm{h}\sim 0.4$~eV) for the two-leg ladder cuprates, we estimate center frequency to be 30~THz and field strength to be 30~MV/cm for our monocycle pump pulse with $E_0=3$. Such a monocycle pulse would be accessible using current laser technology.

\begin{acknowledgments}
This work was supported by the Japan Society for the Promotion of Science, KAKENHI (Grant No.~19H01829, No.~JP19H05825, No.~21H03455, and No.~JP23K13066) from the Ministry of Education, Culture, Sports, Science, and Technology, Japan. Numerical calculation was carried out using computational resources of HOKUSAI at RIKEN Advanced Institute for Computational Science, the supercomputer system at the information initiative center, Hokkaido University, the facilities of the Supercomputer Center at Institute for Solid State Physics, the University of Tokyo, and supercomputer Fugaku provided by the RIKEN Center for Computational Science through the HPCI System Research Project (Project ID: hp220048).
\end{acknowledgments}

\nocite{*}


\begin{thebibliography}{200}
\bibitem{Size} The tDMRG calculation for the $16\times 2$ system using the same pumping condition and $\chi=8000$ density-matrix eigenstates did not give reasonable convergence, because the pumping induces real photoexcition that requires a larger number of $\chi$ for convergence. This is the fsreason why we used tED for the $8\times 2$ system.
\end{thebibliography}

\end{document}